\documentclass[letterpaper,12pt,leqno]{article}
\usepackage{restud,math}
\pdfoutput=1
\hypersetup{pdftitle={u* = √uv}}
\available{https://pascalmichaillat.org/13/}
\newcommand{\bib}{paper.bib}
\newcommand{\pdf}{figures.pdf}

\begin{document}

\title{$\bm{u^{*}=\sqrt{uv}}$}
\author{Pascal Michaillat, Emmanuel Saez
\thanks{Pascal Michaillat: University of California--Santa Cruz. Emmanuel Saez: University of California--Berkeley. We thank Titan Alon, David Baqaee, Daniel Bogart, Gillian Brunet, Varanya Chaubey, Karen Dynan, Gauti Eggertsson, Jim Hamilton, Juan Herreno, Nir Jaimovich, Olivia Lattus, David Lopez-Salido, Magne Mogstad, Emi Nakamura, Edward Nelson, Romain Ranciere, Guillaume Rocheteau, Jon Steinsson, Yuta Takahashi, Naoki Takayama, and Pierre-Olivier Weill for helpful comments and discussions.}}
\date{June 2024}                

\begin{titlepage}\maketitle

This paper aims to compute the unemployment rate $u^*$ that is consistent with full employment in the United States. First, it argues that the most appropriate economic translation of the legal notion of full employment is social efficiency. Here efficiency requires to minimize the nonproductive use of labor---both unemployment and recruiting. The nonproductive use of labor is measured by the number of jobseekers and vacancies, $u + v$. Through the Beveridge curve, the numbers of vacancies and jobseekers are inversely related, $uv = \text{constant}$. With such symmetry the labor market is efficient when there are as many jobseekers as vacancies ($u = v$), inefficiently tight when there are more vacancies than jobseekers ($v > u$), and inefficiently slack when there are more jobseekers than vacancies ($u > v$). Accordingly, the full-employment rate of unemployment (FERU) is the geometric average of the unemployment and vacancy rates: $u^* = \sqrt{uv}$. Between 1930 and 2023, the FERU averages $4.1\%$ and is quite stable---it always remains between $2.5\%$ and $6.6\%$.

\end{titlepage}\section{Introduction}\label{s:introduction}
 
\paragraph{The US government's full-employment mandate} In the United States the federal government and central bank are mandated to maintain the economy at ``full employment,'' or ``maximum employment.'' This legislative mandate comes from the Employment Act of 1946, the Federal Reserve Reform Act of 1977, and the Full Employment and Balanced Growth Act of 1978 \citep{D77,G79,W87a,S11a,B13}. For instance, the Employment Act states that it is the ``policy and responsibility of the federal government\dots to coordinate and utilize all its plans, functions, and resources\dots to promote maximum employment'' \citep[p. 1]{EA}. The Federal Reserve Reform Act of 1977 adds that it is the responsibility of the Federal Reserve ``to promote effectively the goals of maximum employment, stable prices'' \citep[p. 1387]{FRRA}. Finally, the Full Employment and Balanced Growth Act of 1978 was written to ``assert the responsibility of the Federal Government to use all practicable programs and policies to promote full employment'' \citep[p. 1887]{FEBGA}.\footnote{During the debate preceding the Employment Act, maximum employment was considered a less stringent goal than full employment \citep[p. 6]{D77}. In 1978, the Full Employment and Balanced Growth Act amended the Employment Act and replaced maximum employment by the more ambitious target of full employment \citep[p. 398]{W87a}.} In this paper, we aim to compute the unemployment rate that characterizes a state of full or maximum employment. We denote it by $u^*$ and, following \citet{M82a}, we refer to it as the full-employment rate of unemployment (FERU).

\paragraph{Economic translation of the legal notion of full employment} First, we translate the legal notion of full employment into economic terms. Since the Employment Act and Full Employment and Balanced Growth Act clearly state that achieving full employment is a way to maximize social welfare, full employment should be translated to social efficiency. Indeed, the Employment Act states that reaching full employment is designed ``to foster and promote\dots the general welfare''  \citep[p. 1]{EA}. The Full Employment and Balanced Growth Act adds that when the economy departs from full employment,  it ``is deprived of the full supply of goods and services, the full utilization of labor\dots and the related increases in economic well-being that would occur under conditions of genuine full employment'' \citep[p. 1888]{FEBGA}. We therefore measure the FERU by the socially efficient rate of unemployment.

\paragraph{Existing full-employment target: NAIRU} The US government uses two unemployment rates as full-employment target. The first is the non-accelerating-inflation rate of unemployment (NAIRU), which is the unemployment rate at which inflation remains stable, and which is obtained by estimating Phillips curves \citep{SSW97a,G97a,L01,BM02a,CEG19}. For instance, in a recent report, the \citet[p. 2]{JEC19} writes that 
``Today, full employment is considered by many to be synonymous with the non-accelerating inflationary rate of unemployment (NAIRU)---the rate of unemployment that neither stokes nor slows inflation.'' Similarly, the \citet[p. 24]{CEA24} describes the concept of full employment as follows: ``Modern economics has generally defined full employment by citing the theoretical concept of the lowest unemployment rate consistent with stable inflation, which is referred to as $u^*$,\dots the non-accelerating inflationary rate of unemployment (termed NAIRU).'' These quotes are particularly meaningful because they come from the Joint Economic Committee and Council of Economic Advisers, which were both created by the Employment Act of 1946 to ensure that the government achieved its employment mandate. But the NAIRU cannot be the FERU because the NAIRU is not a measure of labor-market efficiency \citep[p. 90]{R97}. The NAIRU might be helpful to achieve the Fed's price-stability mandate, but it says nothing about the government's full-employment mandate.

\paragraph{Existing full-employment target: CBO's NRU} The second target used by the US government is the Congressional Budget Office (CBO)'s noncyclical rate of unemployment (NRU)---which was called ``natural rate of unemployment'' before 2021. The NRU is a slow-moving trend of the unemployment rate computed by aggregating slow-moving changes in the demographic composition of the labor force \citep{B07a,S18a}. For example, when he was President of the Boston Fed, \citet[p. 180]{R14a} measured the departure of the Fed from its full-employment mandate by ``the squared deviations of unemployment from an estimate of full employment utilizing the Congressional Budget Office assessment of the natural rate for each year.'' But trend unemployment is generally not efficient \citep[p. 185]{P00}. Thus, the CBO's NRU cannot be the FERU.

\paragraph{This paper's full-employment target} We compute the FERU as the rate of unemployment that achieves a socially efficient allocation of labor. Such allocation maximizes social output by minimizing the uses of labor that are socially unproductive: both jobseeking and recruiting. The goal is that workers spend as much time as possible producing socially useful things and waste as little time as possible searching for jobs or new hires. Of course, jobseeking and recruiting are necessary for workers and firms to match with each other, but they do not generate any social welfare by themselves.

\paragraph{Social output} The FERU maximizes social output: goods and services produced in the market and at home that engender social welfare. In theory, unemployed workers might produce valuable goods and services at home while looking for jobs. But in practice, the benefits from home production are almost entirely offset by the psychological costs from being unemployed---so the social value of home production is minimal. Furthermore, not all employed workers produce social output. Many workers devote their time to recruiting instead of producing goods and services that add to social welfare. In fact, it takes about one full-time worker to service any vacancy, so the number of recruiters can be counted by the number of vacancies. Accordingly, the share of socially productive workers in the labor force is $1-u-v$, where $u$ is the unemployment rate and $v$ is the vacancy rate. The FERU, therefore, minimizes the sum of the unemployment and vacancy rates, $u + v$.

\paragraph{Beveridge curve} A naive way to minimize $u + v$ would be to set the unemployment rate $u$ and vacancy rate $v$ to zero. But it is impossible to simultaneously reduce the numbers of jobseekers and vacancies because of the Beveridge curve. When the number of jobseekers falls along the Beveridge curve, the number of vacancies necessarily rises; conversely, when the number of vacancies falls, the number of jobseekers necessarily rises. In fact, the Beveridge curve is approximately a rectangular hyperbola:  $uv = A$, where $A>0$ is a constant. Hence, it is infeasible to set the unemployment and vacancy rates to zero, or even to reduce them simultaneously.

\paragraph{Full-employment criterion} Because of the symmetrical roles played by jobseekers and vacancies, the economy is at full employment when there are as many jobseekers as vacancies ($u = v$). This is the only allocation that minimizes $u + v$ subject to $uv = A$. The labor market is inefficiently tight when there are more vacancies than jobseekers ($v > u$), and inefficiently slack when there are more jobseekers than vacancies ($u > v$). This full-employment criterion can be reformulated in terms of labor-market tightness, $v/u$. The tightness gives the number of vacancies per jobseeker. Our analysis implies that the full-employment tightness is $1$. The labor market is inefficiently tight when tightness is above $1$ and inefficiently slack when tightness is below $1$.

\paragraph{Full-employment rate of unemployment} From the Beveridge curve and the equality of the efficient unemployment and vacancy rates, we find that the FERU is the geometric average of the unemployment and vacancy rates: $u^* = \sqrt{uv}$. As it only requires unemployment and vacancy rates, the FERU formula is easy to apply, even in real time. We derived a more general but also more complex formula in \citet{MS21}. Here we show that empirically, the relevant statistics align so that the general formula can be greatly simplified. This provides an incredibly simple, easy to derive, and easy to use formula---which might be especially useful to policymakers.

\paragraph{Deviations from full employment in the United States}  We compute the FERU between 1930 and 2023 in the United States. The FERU averages $4.1\%$. The FERU is also quite stable: it remains between $2.5\%$ and $6.6\%$, while the unemployment rate fluctuates between $1.0\%$ and $25.3\%$. Furthermore, the FERU has generally been below the unemployment rate. That is, the US labor market has generally been inefficiently slack. The unemployment gap $u - u^*$ averages $+2.4$pp. The gap is especially wide in recessions---as wide as $+20.9$pp during the Great Depression and $+5.9$pp during the Great Recession. The US labor market has only been inefficiently tight during major wars---World War 2, Korean War, and Vietnam War---and around the coronavirus pandemic---from 2018Q3 to 2020Q1 and from 2021Q3 to 2023Q4.

\paragraph{Unusuality of the pandemic labor market} As the FERU formula can be applied in real time, we can use it to examine the US labor market during and after the pandemic. We observe that the pandemic labor market has been extremely unusual. First, in 2020, the unemployment gap reached $+6.4$pp. The last time the economy faced such slack was 1940, at the onset of World War 2. Then, in 2022, the unemployment gap bottomed to $-1.5$pp. The last time the economy became so tight was 1945, at the end of World War 2.

\section{Derivation of the FERU formula}\label{s:formula}

Based on the texts of law that introduced the full-employment mandate in the United States, we defined the FERU as the rate of unemployment that achieves a socially efficient allocation of labor. Therefore, the FERU is the solution to the problem of a social planner who allocates labor so as to maximize welfare. We now describe the planner's problem and solve it to derive the FERU formula.

\subsection{Social welfare function}

\paragraph{Maximizing social output} Social output is the production of goods and services that generates social welfare. For simplicity, we exclude distributional considerations from the social welfare function, so social output alone determines social welfare.\footnote{Distributional considerations can be excluded by assuming that workers are risk neutral. If workers are risk averse and are not perfectly insured against unemployment, then the distribution of consumption influences welfare, and the efficient unemployment rate is given by a more complex formula that incorporates distributional elements \citep{LMS10}.} Thus, the social planner allocates labor to maximize social output. This perspective on full employment is consistent with the view expressed by \citet[p. 20]{B44} that ``The material end of all human activity is consumption. Employment is wanted as a means to more consumption\dots as a means to a higher standard of life.'' 

\paragraph{Connection with efficiency in matching models} The concept of efficiency used here is the same as in the modern labor-market models \citep{H90,P00}. These models feature both unemployed workers and job vacancies, each inducing output losses: more unemployment means fewer people at work so less output; more vacancies means more labor devoted to recruiting and also less output. Distributional considerations are typically excluded from the social welfare function \citep[p. 184]{P00}. Hence, the efficient allocation in these models maximizes output by minimizing the output loss caused by unemployment and recruiting.

\subsection{Workers available for market production}

We first determine the pool of labor that is available for market production. 

\paragraph{Full employment among labor-force participants} We assume that the social planner has the entire labor force at its disposal. This assumption follows from the laws that established the full-employment mandate, which were designed to provide employment to labor-force participants. For instance, the Employment Act says that it aims to afford ``useful employment opportunities, including self-employment, for those able, willing, and seeking to work'' \citep[p. 1]{EA}. The Full Employment and Balanced Growth Act uses similar language. Its goal is to ``translate into practical reality the right of all Americans who are able, willing, and seeking to work to full opportunity for useful paid employment'' \citep[p. 1887]{FEBGA}. Thus, the labor force represents the pool of workers that can be tapped for market production. People out of the labor force may be in school or in training, may have retired, or may be looking after their family. They are not available for market production.

\paragraph{Labor force over the business cycle} Although the planner takes the labor force as given, she would have to account for changes in the size of labor force if the size systematically responded to the state of the labor market. In theory, two responses are possible: workers entering the labor force in bad times to supplement their household's income, or jobseekers leaving the labor force in bad times out of discouragement. In practice, however, the labor-force participation rate appears acyclical, so the planner takes the labor-force size as given. Using US data covering 1946--1954, \citet[p. 32]{R57} does not find evidence of the discouraged-jobseeker theory. More systematically, in US data covering 1960--2006, \citet[p. 294]{S09} finds that the labor-force participation rate is acyclical. Similarly, using US data spanning 1976--2009, \citet[pp. 624--625]{RS11} find that over the business cycle, ``the labor force participation rate is nearly constant.'' \citet[p. 19]{EL14} also find that the labor-force participation rate is acyclical in the United States between 1972 and 2007.\footnote{\citet{EL14} argue that high unemployment during the Great Recession caused a drop in labor-force participation. But as \citet{ACF14} and \citet{K17} show, the decline in labor-force participation was primarily caused by population aging and other trends that preceded the Great Recession.} Finally, using a vector autoregression ran on US data for 1976--2016, \citet[figure 1C]{CFM22} find that the impulse response of the labor-force participation rate to a positive productivity shock (a typical shock in business-cycle models) is exactly zero for two years, and while it is slightly positive after two years, it is never significantly different from zero.

\subsection{Social product of employed labor}\label{s:recruiting}

The labor force comprises employed and unemployed workers. We start by assessing the social product of employed labor. 

\paragraph{Breakdown of employed workers' time} We assume that all employed workers have the same productivity. However, these workers are unable to spend their entire time contributing to social output. Instead they must spend some of their time recruiting new hires for their firms. Recruiting takes work: designing and advertising vacancies, screening and interviewing candidates, and negotiating contracts. Beside recruiting, employed workers might also spend time looking for a new job, which takes further time away from socially productive tasks.

\paragraph{National Employer Survey} One source of information about the amount of labor devoted to recruiting in the United States is the National Employer Survey, which was conducted by the Census Bureau in 1997 \citep{V10}. The survey asked thousands of establishments across industries about their recruiting practices \citep{C01a}. Using the survey, \citet[p. 11]{MS21} estimate that the amount of labor required to service a vacancy at any point in time is 0.92 worker. 

\paragraph{Bersin and Associates Survey} A similar number appears in survey data collected by the consulting firm Bersin and Associates in 2011 \citep{GMV18}. The survey asked over 400 firms with more than 100 employees about their spending on all recruiting activities. \citet[p. 2106]{GMV18} find that recruiting one worker costs 0.93 of a monthly wage. Moreover, it takes on average a month to fill a vacancy in the United States \citep[online appendix B]{LMS18}. Combining these two numbers implies that it takes 0.93 worker to service a vacancy.

\paragraph{Labor devoted to recruiting} Overall, both surveys show that it takes about 1 full-time worker to service a vacancy.\footnote{Section~\ref{s:ms21} shows how to extend the FERU formula if the number of recruiters per vacancy is different from 1.} In other words, the number of recruiters in the United States is well measured by the number of vacancies. So the number of workers diverted from producing and allocated to recruiting can be measured by the number of vacancies posted at any point in time.

\paragraph{Labor devoted to job-to-job search} Employed workers might also be distracted from producing if they search for new jobs at work. However, the average number of minutes spent on job search by employed workers is only 31 seconds per day \citep[table 1]{AS20}. So it is a tiny amount taken away from production, which we abstract from.

\subsection{Social product of unemployed labor}

Next, we assess the social product of unemployed labor.

\paragraph{Breakdown of unemployed workers' time} We consider three possible activities for unemployed workers. One, of course, is looking for a job. Jobseeking is required to find employment but---just like recruiting---it does not contribute to social output. Second, unemployed workers produce goods and services at home. Such home production adds to social output and contribute to social welfare. Third, unemployed workers remain idle when they are not looking for jobs and producing at home.

\paragraph{Value of home production versus psychological cost of idleness} The value of jobseekers' home production, net of the psychological cost of idleness, can be estimated from the results by \citet{BM15}. Using administrative data from the US military, \citeauthor{BM15} study how servicemembers choose between reenlisting and leaving the military. The choices allow them to estimate the value of home production plus public benefits minus the psychological cost of idleness during unemployment. Subtracting the value of public benefits from these estimates, \citet[p. 11]{MS21} find that the value of home production minus the psychological cost of idleness relative to the value of market production could be as low as $0.03$. Given such low value, we set the value of home production minus the psychological cost of idleness to zero.\footnote{Section~\ref{s:ms21} shows how to extend the FERU formula if the value of home production minus the psychological cost of idleness is nonzero.} That the social value of unemployment is essentially zero was already noted by \citet[p. 11]{R49}: ``The most important aspect of unemployment is its wastefulness. It is the existence of unused productive resources side by side with unsatisfied human needs that is the intolerable condition.''

\paragraph{Mechanisms behind the psychological cost of idleness} Where do the psychological costs of unemployment come from? The psychological costs associated with unemployment arise from various sources. First, depression, anxiety, and strained personal relations are common consequences of job loss \citep{EL38,T98}.  Job loss is a traumatic event that can lead to a decline in an individual's self-esteem and sense of self-worth \citep{GVD96}. Joblessness also diminishes psychological well-being by creating a sense of helplessness: that one's life is no longer under their control \citep{GD92}. Furthermore, job search appears to reduce unemployed workers' life satisfaction \citep{KM11}. In fact, \citet{J81} emphasizes numerous important benefits of work---which are lost during unemployment. These benefits from work encompass a structured daily routine, regular interactions and shared experiences with individuals beyond the immediate family, the pursuit of overarching goals and purposes, a source of personal status and identity, and the engagement in regular activities. Collectively, the loss of these benefits contribute to the psychological burdens associated with unemployment.

\paragraph{Additional evidence on the psychological cost of idleness} That 
the idleness associated with unemployment can create psychological hardship goes against the idea---standard in neoclassical economics---that unemployed workers enjoy leisure time. Yet, even though it is often neglected in economics, the psychological toll from unemployment has been understood for a long time. \citet[p. 11]{R49} for instance observed that ``The most striking aspect of unemployment is the suffering of the unemployed and their families---the loss of health and morale that follows loss of income and occupation.'' At this point, the detrimental effects of unemployment on mental and physical health are well documented \citep{DFL96,HP00,FS02,W12a,B15a}. A recent field experiment in Bangladesh by \citet{HKL22} illustrates just how large the psychosocial cost of unemployment is. This cost manifests itself in two ways. First, paid employment raises psychosocial well-being substantially more than the same amount of cash alone. Second, two-thirds of employed workers would be willing to forgo cash payments and to continue working for free.

\subsection{Shape of the Beveridge curve}

Given that both unemployed workers and vacant jobs are socially costly, the social planner would want to reduce both. This is not feasible, however, because of the Beveridge curve.

\begin{figure}[p]
\subcaptionbox{Regular scale\label{f:raw1951}}{\includegraphics[scale=\wscale,page=1]{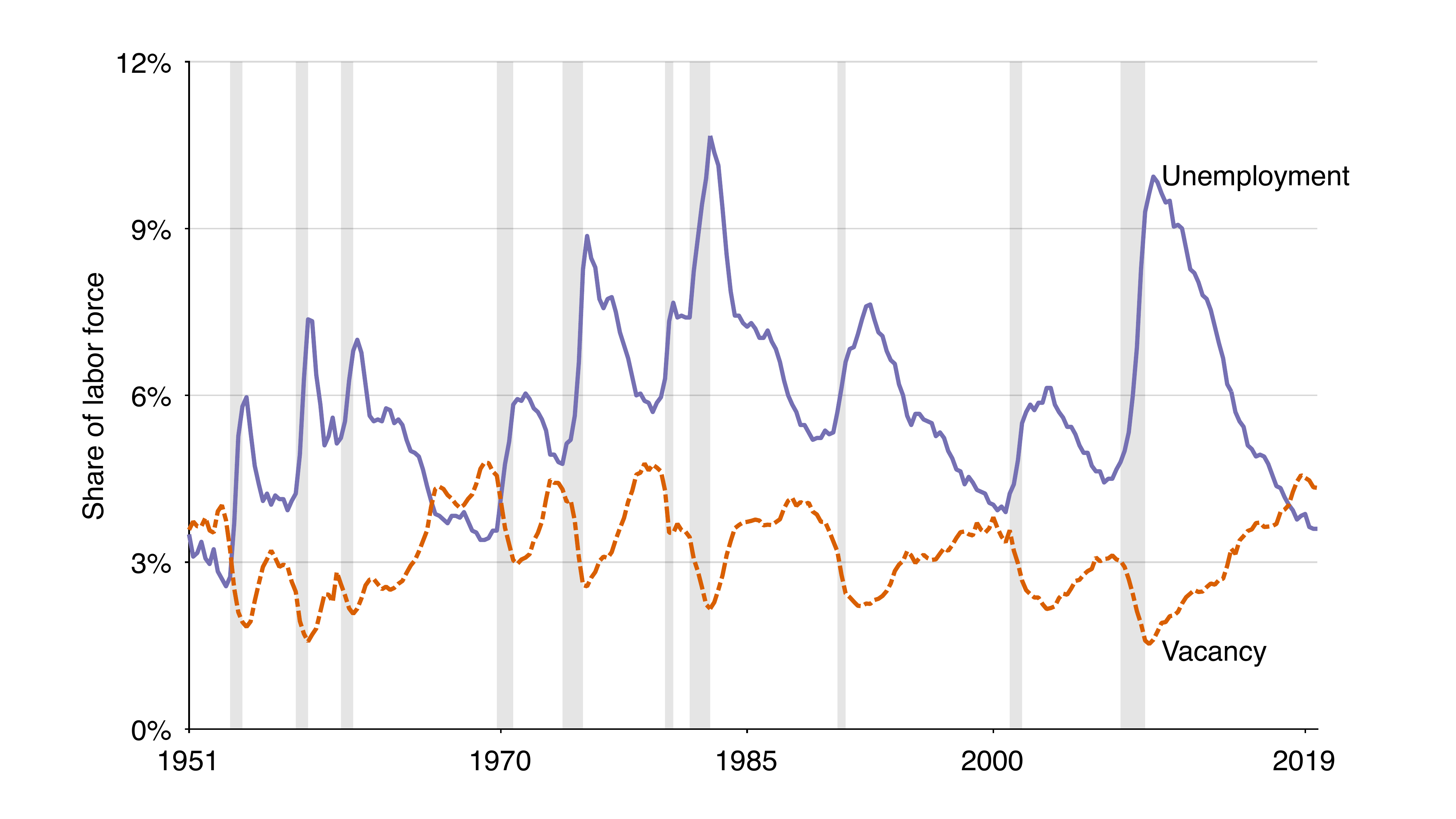}}\\
\subcaptionbox{Log scale\label{f:log1951}}{\includegraphics[scale=\wscale,page=2]{\pdf}}
\caption{Unemployment and vacancy rates in the United States, 1951--2019}
\note{The unemployment rate is measured by the \citet{UNRATE}. Between 1951 and 2000, the vacancy rate is constructed by \citet{B10}; between 2001 and 2019, the vacancy rate is the number of job openings divided by the civilian labor force, both measured by the \citet{CLF16OV,JTSJOL}. Unemployment and vacancy rates are quarterly averages of monthly series. The gray areas are NBER-dated recessions.}
\label{f:data1951}\end{figure}

\paragraph{First observations of the Beveridge curve} Looking at labor-market statistics for Great Britain, \citet{B44} first noted that the number of vacancies and the number of jobseekers move in opposite directions. When the economy is in a slump, there are lots of jobseekers and few vacancies. Conversely, when the economy is in a boom, there are few jobseekers and many vacancies. \citet[figures 1 and 2]{DD58} confirmed this negative relationship between unemployment and vacancies, which has become known as the Beveridge curve, by plotting unemployment and vacancy data for Great Britain, 1946--1956. \citeauthor{DD58}'s first figure is similar to figure~\ref{f:raw1951}. It shows that over time the unemployment rate goes up whenever the vacancy rate goes down, and vice versa. Their second figure is similar to figure~\ref{f:beveridge}. It plots the vacancy rate against the unemployment rate and shows that over the business cycle, unemployment and vacancies move along a downward-sloping curve.

\paragraph{Unemployment and vacancy rates move in opposite directions} The Beveridge curve holds remarkably well in the United States \citep{BD89,EMR15}. Figure~\ref{f:raw1951} depicts the unemployment rate $u$ (number of jobseekers divided by size of the labor force) and the vacancy rate $v$ (number of vacancies divided by size of the labor force) in the United States from 1951 to 2019. The unemployment rate is measured by the \citet{UNRATE}. Between 1951 and 2000, the vacancy rate is constructed by \citet{B10}. Between 2001 and 2019, the vacancy rate is the number of job openings divided by the civilian labor force, both measured by the \citet{CLF16OV,JTSJOL}. The figure shows that unemployment and vacancy rates move in opposite directions. 

\paragraph{Unemployment and vacancy rates are inversely related} In fact, unemployment and vacancy appear to be the inverse of each other, so that doubling the unemployment rate cuts the vacancy rate in half, and conversely, doubling the vacancy rate cuts the unemployment rate in half. Figure \ref{f:log1951} displays again unemployment and vacancy, but now in log scale. The fluctuations of the unemployment and vacancy rates are almost a mirror image of each other, indicating that unemployment and vacancy rates are inversely related.

\paragraph{The Beveridge curve is a rectangular hyperbola} Mathematically, the property that the unemployment rate $u\in [0,1]$ and vacancy rate $v\in [0,1]$ are inversely related implies that the Beveridge curve is a rectangular hyperbola:
\begin{equation*}
vu = A,
\end{equation*}
where $A \in (0,1/4)$ is a constant.\footnote{We impose the condition $A< 1/4$ so the equation $vu = A$ admits at least a solution $(u,v)$ such that $u+v \leq 1$. The condition $u+v\leq 1$ must hold because the number of jobseekers and recruiters is less than the number of labor-force participants. To see where the upper bound $1/4$ comes from, consider the point on the Beveridge curve such that $u=v$. That point satisfies $u^2 = A$ or $u = \sqrt{A}$, and $v = u = \sqrt{A}$. The constraint $u+v\leq 1$ translate into $2 \sqrt{A} \leq 1$, which is equivalent to $A\leq 1/4$. By imposing $A<1/4$, we ensure that parts of the Beveridge curve satisfy the constraint $u+v \leq 1$ (for a reason that will become clear in section~\ref{s:criterion}).}

\paragraph{Estimates of the Beveridge elasticity} We can formally establish that the Beveridge curve is a rectangular hyperbola by estimating the elasticity of the vacancy rate with respect to the unemployment rate, $\oex{v}{u}$. An elasticity of $-1$ corresponds to an hyperbola.  Using the algorithm of \citet{BP98}, and the data displayed in figure~\ref{f:data1951}, \citet[figures 5 and 6]{MS21} estimate the structural breaks of the US Beveridge curve, and the elasticity of the Beveridge curve between these breaks. They find that over the 1951--2019 period, the Beveridge elasticity remains between $-0.84$ and $-1.02$---never far from $-1$. This finding confirms that the US Beveridge curve is close to a rectangular hyperbola.\footnote{Section~\ref{s:ms21} shows how to extend the FERU formula if the Beveridge curve is an isoelastic curve with an elasticity different from 1.}

\paragraph{Foundation for the hyperbolic Beveridge curve} It is quite natural that the empirical Beveridge curve is a rectangular hyperbola, since this is the shape that arises in the most basic matching model of the labor market. In matching models, the Beveridge curve is the locus of points such that labor-market flows are balanced: the number of workers who lose or quit their jobs equals the number of workers who find a job. The employment rate $1-u$ is approximately constant at 1 since the unemployment rate $u$ is an order of magnitude less than 1. The job-separation rate $\l$ is also constant, so the number of job separations $\l \times (1-u)$ is approximately constant at $\l$. So along the Beveridge curve, the number of workers who find a job is constant at $\l$. With the standard symmetric Cobb-Douglas matching function, $m = \o \times \sqrt{uv}$, the number of workers who find a job at any point in time is proportional to $\sqrt{uv}$.\footnote{The US matching function appears to have a Cobb-Douglas form with exponents of $0.5$ on unemployment and vacancies \citep[p. 9]{MS21}.} Hence, along the Beveridge curve, $\sqrt{uv}$ and thus $uv$ must be constant: the Beveridge curve is a rectangular hyperbola.

\paragraph{Job-to-job transitions and labor-force transitions} We have just provided a foundation for the hyperbolic Beveridge curve based on a basic matching model. But our analysis is in no way limited to such model. It only presumes that the Beveridge curve exists, but does not put additional restrictions on the structure of the labor market. For instance, the basic matching model only features flows between employment and unemployment. In practice, there are vast flows from employment to employment and in and out of the labor force \citep[figure 1]{BD90}. Our analysis is in the sufficient-statistic tradition \citep{C09}: it applies to all models with such flows as long as they feature a Beveridge curve and constant labor force. The Beveridge curve summarizes everything we need to know for the welfare analysis.\footnote{An implicit assumption is that all workers have the same productivity in all firms. Thus, job-to-job and labor-force transitions do not affect the output of transiting workers, and do not affect welfare. (Since the size of the labor force is constant, a worker leaving the labor force must be replaced by a new worker entering the labor force; for instance, a worker going on parental leave is replaced by a worker finishing parental leave.)}

\paragraph{Unemployment dynamics} The matching foundation for the Beveridge curve  considers a state in which labor-market flows are balanced. But unemployment perpetually evolves through a dynamic process driven by differences between inflows into unemployment (job separation) and outflows from unemployment (job finding). The unemployment rate is on the Beveridge curve only when inflows and outflows are balanced. However, as \citet[p. 236]{P09a} observes, ``Perhaps surprisingly at first, but on reflection not so surprisingly, we get a good approximation to the dynamics of unemployment if we treat unemployment as if it were always on the Beveridge curve.'' The reasons is that in the United States, inflows and outflows are extremely large, so the dynamic unemployment process converges extremely quickly to the Beveridge curve. Formally, \citet[p. 7]{MS21} show that 50\% of the deviation of the unemployment rate from the Beveridge curve evaporates within one month, and 90\% evaporates within one quarter. Thus, the US unemployment rate is always close to the Beveridge curve. This explains why many matching models assume that the Beveridge curve holds at all times \citep{H05,P09}.

\subsection{Full-employment criterion}\label{s:criterion}

Using the social product of employed and unemployed labor and the shape of the Beveridge curve, we now formally describe and solve the social planner's problem.

\paragraph{Objective} The planner aims to maximize social output. Since the size of the labor force is fixed and unemployment is completely wasteful, the objective is to minimize the amount of labor in unemployment or recruiting. And since the number of recruiters can be counted by the number of vacancies, the objective is to minimize the number of unemployed workers and vacancies. Equivalently, the labor-force size being fixed, the objective is to minimize the sum of the unemployment and vacancy rates, $u + v$. 

\paragraph{Optimization constraint} This minimization is subject to the Beveridge curve constraint, $u v = A$. Because of the Beveridge curve, it is not possible to reduce unemployment and vacancies at the same time, so the planner must trade off unemployment and vacancies. The planner takes the Beveridge curve as given because the Beveridge curve does not seem to respond to monetary or fiscal stabilization policy. Indeed, in many business-cycle models with unemployment, the Beveridge curve is unaffected by monetary and fiscal policy \citep{BG10,RW11,M14,MS19,MS22,MS24}. In these models the Beveridge curve is determined by the matching function and job-separation rate. Neither responds to monetary or fiscal policy, so the Beveridge curve is unaffected by policy.\footnote{Other policies do affect the Beveridge curve. For instance, a reduction in unemployment insurance bolsters jobseekers' search effort, which shifts the Beveridge curve inward \citep{LMS18,HKM21}. Structural policies designed to improve labor-market flows, such as the German Hartz reforms of 2003--2005, might also shift the Beveridge curve inward \citep{FS09,KW16}.} 

\paragraph{Optimization problem}  The planner minimizes nonproduction $u + v$ subject to the Beveridge curve $uv = A$, with $u \in [0,1]$ and $v\in [0,1]$. To simplify the problem, we substitute the Beveridge curve, $v = A/u$, into the objective function. Then the problem simply is to minimize $u+A/u$ over $u \in [A,1]$.\footnote{With $u \in [A,1]$, we ensure that $v = A/u$ is in $[0,1]$. In fact, $v \in [A,1]$, just like $u$.} The function $u \mapsto u+A/u$, with domain $[A,1]$, is continuous and strictly convex.\footnote{To see that the function is strictly convex, note that its second derivative is positive: $2A/u^3>0$.} Therefore, the function admits a unique minimum on $[A,1]$.\footnote{Note that at $u=A$ and $u=1$ the value of the function is that same, $u+A/u = 1+A$. So neither $u=A$ nor $u=1$ can be the minimum point---otherwise the minimum would not be unique. Thus, the function's minimum sits on ($A,1)$.}

\paragraph{Solution by symmetry} The planner minimizes $u+v$ subject to $u v = A$, with $u\in [0,1]$ and $v\in [0,1]$. As we have just seen, this minimization problem admits a unique minimum. Since the problem is perfectly symmetric in $u$ and $v$, the minimum must be reached when $u = v$. Therefore, full employment is reached when the unemployment and vacancy rates are equal.

\paragraph{Deviation from full employment} When the unemployment and vacancy rates are not equal, the labor market is operating inefficiently. The labor market is inefficiently tight when there are more vacancies than jobseekers ($v > u$). In that case, increasing $u$ and reducing $v$ would increase social output. The labor market is inefficiently slack when there are more jobseekers than vacancies ($u > v$). Then, reducing $u$ and increasing $v$ would increase social output.

\paragraph{First-order condition} We can also find the minimum of $u+A/u$ over $u \in (A,1)$ by first-order condition. Since $u \mapsto u+A/u$ is strictly convex, the first-order condition is necessary and sufficient to find the minimum. We take the derivative of $u+A/u$ with respect to $u$ and set it to $0$. We obtain $1- A/u^2 = 0$. This condition implies that the minimum occurs when $u = \sqrt{A}$. By the Beveridge curve we have $v = A/u$, so at the minimum $v = A/\sqrt{A} = \sqrt{A}$. Accordingly, the full-employment unemployment and vacancy rates satisfy 
\begin{equation}
u^* = v^* = \sqrt{A}.
\label{e:location}\end{equation}
Once again, we find that at full employment, the unemployment and vacancy rates are equal. Equation \eqref{e:location} also shows that the location of Beveridge curve solely determines the unemployment and vacancy rates at full employment.\footnote{Because the number of jobseekers and recruiters cannot exceed the number of labor-force participants, the planner's problem also includes the constraint $u+v \leq 1$. But the constraint is satisfied at the optimum, so it does not alter the solution to the planner's problem. Indeed, we have $A <1/4$, so $\sqrt{A} < 1/2$, which implies that $u^* + v^* = 2 \times \sqrt{A} < 1$.}

\paragraph{Tightness formulation} The full-employment criterion can be reformulated in terms of labor-market tightness, $v/u$. The tightness gives the number of vacancies per jobseekers. The analysis implies that the labor market is at full employment when tightness is 1, is inefficiently tight when tightness is above 1, and is inefficiently slack when tightness is below 1. 

\subsection{FERU}

\paragraph{Formula}  We have seen that the FERU is given by $u^* = \sqrt{A}$ where the parameter $A$ determines the location of Beveridge curve, $uv = A$. Hence, the FERU is the geometric average of the unemployment and vacancy rates: 
\begin{equation}
u^* = \sqrt{uv}. 
\label{e:feru}\end{equation}
Since $uv = A>0$, one basic implication of \eqref{e:feru} is that $u^*>0$. In other words, the FERU is not zero: reaching full employment should not be interpreted as reaching zero unemployment.

\paragraph{The FERU is not zero, because zero unemployment is not feasible} The FERU is not zero because the Beveridge curve prevents unemployment from reaching zero. Because each vacancy requires a recruiter, the vacancy rate $v$ is at most 1. Accordingly, the Beveridge curve $u = A/v$ prevents the unemployment rate to fall below $A>0$. The fact that labor-market flows impose a minimum level of unemployment---and therefore that full employment cannot be zero unemployment---has been known for a long time. \citet[p.~125]{B44} realized that ``However great the unsatisfied demand for labor, there is an irreducible minimum of unemployment, a margin in the labor force required to make change and movement possible.'' As a result, ``even under full employment, there will be some unemployment,\dots on each day some men able and willing to work will not be working.'' Similarly, \citet[pp. 169--170]{R46} noted that ``In a changing world there are always bound to be, at any moment, some workers who have left one job and have not yet found another.\dots Changes in occupation for personal reasons will always be going on. So long as such shifts in employment are taking place there is always likely to be some unemployment even when the general demand for labor is very high. Thus completely full employment can never be seen.''

\paragraph{The FERU is not zero, because zero unemployment is not desirable} Second, the FERU is not zero because it is never efficient to reduce the unemployment rate to zero. Unemployment is clearly a waste of economic resources as people who would like to work are not able to be productive. Yet, reducing the unemployment rate to zero is not desirable because it would require to divert a vast amount of labor toward recruiting. In fact, it is not efficient to reduce the unemployment rate below the vacancy rate. Reducing the unemployment rate by 1\% requires to raise the vacancy rate by 1\%, due to the hyperbolic Beveridge curve. When the unemployment rate is less than the vacancy rate, the increase in vacancy rate is more than the decrease in unemployment rate. Hence, overall, although the unemployment rate falls, the sum of the unemployment and vacancy rates increases---which means that social output falls.\footnote{Zero unemployment is not desirable here because of the resources absorbed by recruiting. \citet[p. 170]{R46} agreed that ``no-one regards 100\% employment as a desirable objective.'' Her logic was different, however. She argued that ``the attainment of full employment, in this absolute sense, would require strict controls, including direction of labor'' and that it would ``involve great sacrifices of liberty,'' even the ``complete conscription of labor.''}

\section{Deviation from full employment in the United States}\label{s:us}

We compute the FERU in the United States over three periods: the standard postwar period, 1951--2019; the Great Depression and World War 2, 1930--1950; and the coronavirus pandemic, 2020--2023.

\subsection{Postwar period}\label{s:postwar}

We first focus on the postwar period, 1951--2019. This is a standard period in the macro-labor literature, for which the unemployment and vacancy data are well known and well understood. For instance, \citet{S05}'s seminal paper starts in 1951. We stop at the end of 2019 to avoid incorporating the pandemic, which is an extremely unusual period, and which we will discuss later.

\paragraph{Unemployment and vacancy rates} We use the standard, official measure of unemployment rate, constructed by the \citet{UNRATE} from the Current Population Survey (CPS).\footnote{The measure used here is labelled U3 by the \citet{BLS23}; it only includes jobseekers who want a job, are available to start a job, and have been actively searching for a job in the past 4 weeks. Section~\ref{s:u4u5} repeats the analysis with two broader measures of unemployment that include jobseekers with lower search effort: U4 and U5. These measures add to U3 workers who want a job, are available to start a job, have been actively searching for a job in the past 12 months but not in the past 4 weeks.} We measure the vacancy rate from two different sources, because there is no continuous vacancy series over the period. For 1951--2000, we use the vacancy rate constructed by \citet{B10}. This series is based on the help-wanted advertising index constructed by the Conference Board, corrected to account for the shift from print advertising to online advertising in the 1990s. The Conference Board index was compiled by aggregating help-wanted advertising in major metropolitan newspapers in the United States. It has been showed to be a good proxy for job vacancies, and it has become standard in the macro-labor literature \citep{A87,S05}. For 2001--2019, we use the number of job openings measured by the \citet{JTSJOL} from the Job Opening and Labor Turnover Survey (JOLTS), divided by the civilian labor force constructed by the \citet{CLF16OV} from the CPS.\footnote{To best align unemployment and vacancy data, we shift forward by one month the number of job openings from JOLTS. For instance, we assign to December 2023 the number of job openings that the BLS assigns to November 2023. The motivation for this shift is that the number of job openings refers to the last business day of the month (for instance, Thursday 30 November, 2023), while the unemployment rate from CPS refers to the Sunday--Saturday week including the 12th of the month (for instance, Sunday 10 December 2023 to Saturday 16 December 2023) \citep{BLS20,BLS24}. So in any given month, the number of job openings refers to a day that is closer to next month's CPS reference week than to this month's CPS reference week.} We then splice the two series to obtain the vacancy rate for 1951--2019. The unemployment rate, $u$, and vacancy rate, $v$ are plotted in figure~\ref{f:data1951}.

\begin{figure}[p]
\subcaptionbox{Visualization based on unemployment and vacancy rates\label{f:uv1951}}{\includegraphics[scale=\wscale,page=3]{\pdf}}\\
\subcaptionbox{Visualization based on labor-market tightness\label{f:tightness1951}}{\includegraphics[scale=\wscale,page=4]{\pdf}}
\caption{Deviation from full employment in the United States, 1951--2019}
\note{Unemployment and vacancy rates come from figure~\ref{f:data1951}. Labor-market tightness is the ratio of the vacancy and unemployment rates. The gray areas are NBER-dated recessions. The labor market is at full employment when the unemployment rate equals the vacancy rate, is inefficiently slack when the unemployment rate is above the vacancy rate (purple shade), and is inefficiently tight when the unemployment rate is below the vacancy rate (orange shade). The US labor market is inefficiently tight during the Korean War (1951Q1--1953Q3) and Vietnam War (1965Q4--1970Q1), and at the end of the Trump presidency (2018Q2--2019Q4); it is inefficiently slack during the rest of the period 1951--2019.}
\label{f:state1951}\end{figure}

\begin{figure}[p]
\subcaptionbox{Construction of the FERU\label{f:ustar1951}}{\includegraphics[scale=\wscale,page=5]{\pdf}}\\
\subcaptionbox{Distance from the FERU\label{f:gap1951}}{\includegraphics[scale=\wscale,page=6]{\pdf}}
\caption{FERU in the United States, 1951--2019}
\note{Unemployment rate $u$ and vacancy rate $v$ come from figure~\ref{f:data1951}. The FERU is $u^* = \sqrt{uv}$. The gray areas are NBER-dated recessions.}
\label{f:feru1951}\end{figure}

\paragraph{Deviation from full employment} We assess the state of the US labor market between 1951 and 2019. We use the unemployment and vacancy rates from figure~\ref{f:data1951}. The labor market is inefficiently slack whenever the unemployment rate is above the vacancy rate; it is inefficiently tight whenever the unemployment rate is below the vacancy rate. The unemployment rate averages $5.8\%$ over the period, while the vacancy rate only averages $3.2\%$. So on average, the unemployment rate is markedly higher than the vacancy rate, which shows that the labor market is inefficiently slack.
In fact, between 1951 and 2019, the labor market is always inefficiently slack except in three episodes (figure~\ref{f:uv1951}): the Korean War (1951Q1--1953Q3), the Vietnam War (1965Q4--1970Q1), and the end of the Trump presidency (2018Q2--2019Q4). 

\paragraph{Visualization with labor-market tightness} The state of the US labor market can also be visualized by plotting the labor-market tightness $v/u$ (figure~\ref{f:tightness1951}). The labor market is inefficiently slack whenever tightness is below 1, inefficiently tight whenever tightness is above 1, and at full employment when tightness equals 1---when there is just one vacancy per jobseeker. Tightness averages $0.62$ between 1951 and 2019, well below 1, which is another manifestation that the labor market is inefficiently slack on average. Tightness peaked at $1.49$ in 1953Q1, during the Korean War, and it bottomed at $0.16$ in 2009Q3, during the Great Recession.

\paragraph{Full-employment unemployment rate} We then compute the FERU between 1951 and 2019 using the formula $u^* = \sqrt{uv}$. The FERU is fairly stable: it remains between $3.0\%$ and $5.3\%$, with an average value of $4.2\%$ (figure~\ref{f:ustar1951}). The Beveridge curve shifts in and out during the postwar period \citep[figure 1]{MS21}, but the shifts are not large enough to produce noteworthy changes in the FERU.

\paragraph{Unemployment gap} We also compute the unemployment gap, which is the difference $u-u^*$ between actual unemployment rate and FERU. The unemployment gap indicates the distance from full employment at any given time. The unemployment gap is generally positive and sharply countercyclical (figure~\ref{f:gap1951}). The unemployment gap averages $1.6$pp over the 1951--2019 period. The gap peaked at $+5.9$pp in 2009Q4, during the Great Recession; it reached the same value in 1982Q4, at the end of the Volcker recession. By contrast, the lowest value taken by the unemployment gap is $-0.6$pp. The gap reached this value twice: in 1953Q1, during the Korean War; and in 1969Q1, during the Vietnam War. Hence, the economy is generally not at full employment, and it is especially far from full employment in recessions.

\subsection{Great Depression and World War 2}\label{s:depression}

Next, we apply our full-employment criterion and FERU formula to the historical period 1930--1950, which covers both the Great Depression and World War 2. Due to its simplicity, the FERU formula can easily be applied to historical data.

\paragraph{Unemployment and vacancy rates} The unemployment and vacancy rates for 1930--1950 are constructed by \citet{PZ21}. The vacancy rate is based on the Metropolitan Life Insurance Company help-wanted index, which was compiled by aggregating help-wanted advertising in newspapers across major US cities, and appears to be a good proxy for job vacancies \citep{Z98a}. The unemployment rate is constructed from the annual unemployment series computed by \citet{W92a}, which is extended to a monthly series using unemployment rates compiled by the National Bureau of Economic Research (NBER) from various sources.
Between 1930 and 1950 it remains true that unemployment and vacancy rates move in opposite directions (figure~\ref{f:raw1930}).\footnote{\citet{PZ21} produce a vacancy series that starts in 1919 and an unemployment series that starts in 1890. \citet[p. 339]{Z98a} argues, however, that the vacancy numbers are unreliable for 1919--1923, because some important newspaper data are missing during that time. Moreover, there is no monthly measure of unemployment between 1890 and 1929. Instead, the monthly unemployment fluctuations are inferred from the spread between the yields of bonds of different quality. Given these limitations, we begin our analysis in 1930.} 

\begin{figure}[p]
\subcaptionbox{Regular scale\label{f:raw1930}}{\includegraphics[scale=\wscale,page=8]{\pdf}}\\
\subcaptionbox{Log scale\label{f:log1930}}{\includegraphics[scale=\wscale,page=9]{\pdf}}
\caption{Unemployment and vacancy rates in the United States, 1930--1950}
\note{The unemployment and vacancy rates are constructed by \citet{PZ21}. Unemployment and vacancy rates are quarterly averages of monthly series. The gray areas are NBER-dated recessions.}
\label{f:data1930}\end{figure}

\paragraph{Hyperbolic Beveridge curve} Using a log scale, it also appears that unemployment and vacancy rates are inversely related (figure~\ref{f:log1930}). These fluctuations indicate that just as in the postwar era, the Beveridge curve is close to a rectangular hyperbola in 1930--1950. We also compute the elasticity of the 1930--1950 Beveridge curve by running an OLS regression of log vacancy rate on log unemployment rate. We find an elasticity of $-0.79$, which is not far from the elasticity of $-1$ for a rectangular hyperbola, and is close to the elasticity of $-0.84$ for the 1951--1961 Beveridge curve \citep[figure 6]{MS21}. The 1930--1950 period saw vast fluctuations in unemployment and vacancy rates: the unemployment rate fluctuated between $1.0\%$ and $25.3\%$; the vacancy rate fluctuated between $0.7\%$ and $6.7\%$. Yet the hyperbolic shape of the Beveridge curve held well. 

\paragraph{Deviation from full employment} The unemployment rate averages $9.0\%$ over the period, while the vacancy rate only averages $2.3\%$. So on average the unemployment rate is markedly higher than the vacancy rate, which indicates that the US labor market is inefficiently slack on average. In fact, the US labor market is always inefficiently slack between 1930 and 1950 except at the end World War 2, during 1942Q3--1946Q3, when it was inefficiently tight (figure~\ref{f:uv1930}).

\begin{figure}[p]
\subcaptionbox{Visualization based on unemployment and vacancy rates\label{f:uv1930}}{\includegraphics[scale=\wscale,page=10]{\pdf}}\\
\subcaptionbox{Visualization based on labor-market tightness\label{f:tightness1930}}{\includegraphics[scale=\wscale,page=11]{\pdf}}
\caption{Deviation from full employment in the United States, 1930--1950}
\note{Unemployment and vacancy rates come from figure~\ref{f:data1930}. Labor-market tightness is the ratio of the vacancy and unemployment rates. The gray areas are NBER-dated recessions. The labor market is at full employment when the unemployment rate equals the vacancy rate, is inefficiently slack when the unemployment rate is above the vacancy rate (purple shade), and is inefficiently tight when the unemployment rate is below the vacancy rate (orange shade). The US labor market is inefficiently tight during World War 2 (1942Q3--1946Q3); it is inefficiently slack during the rest of the period 1930--1950.}
\label{f:state1930}\end{figure}

\begin{figure}[p]
\subcaptionbox{Construction of the FERU\label{f:ustar1930}}{\includegraphics[scale=\wscale,page=12]{\pdf}}\\
\subcaptionbox{Distance from the FERU\label{f:gap1930}}{\includegraphics[scale=\wscale,page=13]{\pdf}}
\caption{FERU in the United States, 1930--1950}
\note{Unemployment rate $u$ and vacancy rate $v$ come from figure~\ref{f:data1930}. The FERU is $u^* = \sqrt{uv}$. The gray areas are NBER-dated recessions.}
\label{f:feru1930}\end{figure}

\paragraph{Visualization with labor-market tightness} The state of the US labor market can also be visualized by plotting labor-market tightness (figure~\ref{f:tightness1930}). 
Tightness averages $0.85<1$ between 1930 and 1950, which confirms that the US labor market is inefficiently slack on average. Tightness is extremely volatile during the period: it plunged to $0.03$ in 1932Q3, during the Great Depression, and climbed all the way to $6.8$ in 1944Q4, at the end of World War 2. 

\paragraph{Full-employment unemployment rate} We then compute the FERU between 1930 and 1950 using the formula $u^* = \sqrt{uv}$. Despite the period's macroeconomic volatility, the FERU is quite stable: it stays between $2.5\%$ and $4.6\%$, with an average value of $3.5\%$ (figure~\ref{f:ustar1930}).

\paragraph{Unemployment gap} Finally, we compute the unemployment gap $u-u^*$. The unemployment gap averages $+5.5$pp between 1930 and 1950 (figure~\ref{f:gap1930}). The unemployment gap was of course positive and very large during the Great Depression: the labor market was much too slack then. The unemployment gap reached $+20.9$pp in 1932Q3. The economy recovered only slowly from the depression. The economy only reached full employment in 1942Q3, a few quarters after the United States had entered World War 2. The unemployment gap kept falling during the war; it reached $-1.6$pp in 1945Q1. The unemployment gap turned positive again during the 1948--1949 recession.

\subsection{Coronavirus pandemic}\label{s:pandemic}

Finally, we apply our full-employment criterion and FERU formula to the coronavirus pandemic in the United States. We focus on the three-year period between January 2020 and December 2023. Here, the simplicity of our FERU formula allows us to apply to real-time data and assess the current state of the labor market.

\paragraph{Unemployment and vacancy rates} The unemployment rate is measured by the \citet{UNRATE} from the CPS. The vacancy rate is the number of job openings measured by the  \citet{JTSJOL} from the JOLTS, divided by the civilian labor force measured by the \citet{CLF16OV} from the CPS. Both series are displayed on figure~\ref{f:uv2020}. The unemployment rate averages $5.2\%$ over 2020--2023. The vacancy rate averages $5.6\%$ over the same period.

\begin{figure}[p]
\subcaptionbox{Visualization based on unemployment and vacancy rates\label{f:uv2020}}{\includegraphics[scale=\wscale,page=14]{\pdf}}\\
\subcaptionbox{Visualization based on labor-market tightness\label{f:tightness2020}}{\includegraphics[scale=\wscale,page=15]{\pdf}}
\caption{Deviation from full employment in the United States, 2020--2023}
\note{The unemployment rate is measured by the \citet{UNRATE}. The vacancy rate is the number of job openings divided by the civilian labor force, both measured by the \citet{CLF16OV,JTSJOL}. Unemployment and vacancy rates are quarterly averages of monthly series. The labor-market tightness is the ratio of the vacancy and unemployment rates. The gray area is the NBER-dated pandemic recession. The labor market is at full employment when the unemployment rate equals the vacancy rate, is inefficiently slack when the unemployment rate is above the vacancy rate (purple shade), and is inefficiently tight when the unemployment rate is below the vacancy rate (orange shade). The US labor market is inefficiently tight in 2021Q3--2023Q4 and inefficiently slack in 2020Q2--2021Q2.}
\label{f:state2020}\end{figure}

\begin{figure}[p]
\subcaptionbox{Construction of the FERU\label{f:ustar2020}}{\includegraphics[scale=\wscale,page=16]{\pdf}}\\
\subcaptionbox{Distance from the FERU\label{f:gap2020}}{\includegraphics[scale=\wscale,page=17]{\pdf}}
\caption{FERU in the United States, 2020--2023}
\note{Unemployment rate $u$ and vacancy rate $v$ come from figure~\ref{f:uv2020}. The FERU is $u^* = \sqrt{uv}$. The gray area is the NBER-dated pandemic recession.}
\label{f:feru2020}\end{figure}

\paragraph{Deviation from full employment} The labor market is inefficiently slack whenever the unemployment rate $u$ is above the vacancy rate $v$; it is inefficiently tight whenever the unemployment rate is below the vacancy rate. We find that the US labor market is inefficiently slack from 2020Q2 to 2021Q2; it is inefficiently tight from 2021Q3 to 2023Q4 (figure~\ref{f:uv2020}).

\paragraph{Visualization with labor-market tightness} The state of the US labor market can also be visualized by plotting labor-market tightness $v/u$ (figure~\ref{f:tightness2020}). Tightness averages $1.31$ over the 2020--2023 period. Tightness cratered to $0.26$ in 2020Q2, so the labor market was much too slack at the beginning of the pandemic. The labor market then recovered and passed the point of full employment ($v/u=1$) in the middle of 2021. The labor market then steadily tightened to reach $1.97$ in 2022Q2. At that point, the labor market was much too tight. After peaking in 2022Q2, tightness slowly fell to reach $1.44$ at the end of 2023. So the labor market remains inefficiently tight at the end of 2023, but it is nearing full employment.

\begin{figure}[t]
\includegraphics[scale=\wscale,page=18]{\pdf}
\caption{Beveridge curve in the United States, 2001--2023}
\note{Each dot represents the unemployment and vacancy rates in a quarter between 2001 and 2023. The unemployment rate is measured by the \citet{UNRATE}. The vacancy rate is the number of job openings divided by the civilian labor force, both measured by the \citet{CLF16OV,JTSJOL}. The labor market is at full employment when the unemployment rate equals the vacancy rate (pink line); it is inefficiently slack when the vacancy rate is below the unemployment rate (purple dots); it is inefficiently tight when the vacancy rate is above the unemployment rate (orange dots).}
\label{f:beveridge}\end{figure}

\paragraph{Full-employment unemployment rate} Between 2020 and 2023, the FERU averages $5.1\%$ (figure~\ref{f:ustar2020}). The FERU was $4.0\%$ in 2020Q1, at the onset of the pandemic, but it sharply increased to $6.6\%$ in the next quarter. It hovered around $6.0\%$ during the rest of 2020--2021, and slowly decreased to $4.5\%$ at the end of 2023.

\paragraph{Unemployment gap} We also compute the unemployment gap $u-u^*$ (figure~\ref{f:gap2020}). While the unemployment gap averages 0 over the period, the labor market experienced sharp departures from full employment. The unemployment gap was initially positive and large: the labor market was much too slack in the first year of the pandemic. The unemployment gap reached $+6.4$pp in 2020Q2. But the economy recovered quickly and reached full employment in the middle of 2021. The unemployment gap turned negative after that, reaching $-1.5$pp in 2022Q2. The gap then shrunk to $-0.7$pp in 2023Q4. So in 2022--2023, the labor market was well beyond full employment.

\paragraph{The sudden rise of the FERU} The FERU increased by more than 2pp at the onset of the pandemic. Such a sharp increase is unprecedented. It can be explained by the gigantic outward shift of the Beveridge curve that took place in the spring 2020. Graphically, the FERU appears at the intersection of the Beveridge curve and the identity line (figure~\ref{f:beveridge}). In 2020Q1, at the onset of the pandemic, the labor market was close to full employment, and the unemployment rate was at $3.8\%$. In 2021Q2, about a year later, the labor market had returned to the vicinity of full employment, but the unemployment rate was now $5.9\%$. Indeed, the FERU rose from $4.0\%$ to $5.8\%$ from 2020Q1 to 2021Q2. This rise was caused by the outward shift of the Beveridge curve that occurred between the first and second quarters of 2020. Mathematically, the FERU is determined by the location of the Beveridge curve (equation~\eqref{e:location}), so only a sharp outward shift of the Beveridge curve can increase the FERU.

\subsection{Complete 1930--2023 period}\label{s:complete}

To conclude, we combine the US unemployment and vacancy rates from 1930 to 2023.

\begin{figure}[p]
\subcaptionbox{Visualization based on unemployment and vacancy rates\label{f:uv}}{\includegraphics[ scale=\wscale,page=19]{\pdf}}\\
\subcaptionbox{Visualization based on labor-market tightness\label{f:tightness}}{\includegraphics[ scale=\wscale,page=20]{\pdf}}
\caption{Deviation from full employment in the United States, 1930--2023}
\note{Unemployment and vacancy rates are obtained by splicing the unemployment and vacancy rates from figures~\ref{f:data1951}, \ref{f:data1930}, and \ref{f:uv2020}. Labor-market tightness is the ratio of the vacancy and unemployment rates. The gray areas are NBER-dated recessions. The labor market is at full employment when the unemployment rate equals the vacancy rate, is inefficiently slack when the unemployment rate is above the vacancy rate (purple shade), and is inefficiently tight when the unemployment rate is below the vacancy rate (orange shade). The figure summarizes the findings from figures \ref{f:state1951}, \ref{f:state1930}, and \ref{f:state2020}.}
\label{f:state}\end{figure}

\paragraph{Deviation from full employment} Between 1930 and 2023, the unemployment rate averages $6.5\%$. The vacancy rate averages slightly less than half of that, $3.1\%$. The unemployment rate is generally above the vacancy rate, and this gap is exacerbated in recessions (figure~\ref{f:uv}). This means that the labor market does not generally operate at full employment; instead, it is generally inefficiently slack. Before 2018, the labor market had only been inefficiently tight during major wars---World War 2, Korean War, Vietnam War. Since then the labor market has been inefficiently tight just before and after the coronavirus pandemic: from 2018Q3 to 2020Q1 and from 2021Q3 to 2023Q4. The state of the labor market around the pandemic is a rarity. It is the only peacetime episode of inefficiently tight labor market. \citet[p. 322]{K36} doubted that the labor market could reach full employment in peacetime. He was essentially right: before 2018 the US labor market had only reached full employment during wartime.

\begin{figure}[p]
\subcaptionbox{Value of the FERU\label{f:ustar}}{\includegraphics[scale=\wscale,page=21]{\pdf}}\\
\subcaptionbox{Distance from the FERU\label{f:gap}}{\includegraphics[scale=\wscale,page=22]{\pdf}}
\caption{FERU in the United States, 1930--2023}
\note{The FERU is obtained by splicing the FERUs from figures~\ref{f:ustar1951}, \ref{f:ustar1930}, and \ref{f:ustar2020}. The unemployment rate is obtained by splicing the unemployment rates from figures~\ref{f:data1951}, \ref{f:data1930}, and \ref{f:uv2020}.  The gray areas are NBER-dated recessions.}
\label{f:feru}\end{figure}

\paragraph{Visualization with labor-market tightness} The state of the labor market can also be visualized by plotting labor-market tightness (figure~\ref{f:tightness}). Over 1930--2023, labor-market tightness averages $0.69$. Tightness is extremely volatile before the end of World War 2. Tightness records its most extreme fluctuations during that period: tightness plunged to $0.03$ in 1932Q3, during the Great Depression, and climbed all the way to $6.8$ in 1944Q4, at the end of World War 2. In 2022, tightness reached a value of $1.97$, which it had last reached in 1945. In the recovery of the pandemic, the US labor market has become historically tight.

\paragraph{FERU} Over 1930--2023, the FERU averages $4.1\%$ (figure~\ref{f:ustar}). The FERU is stable over time, remaining between $2.5\%$ and $6.6\%$ over almost a century. It hovered around $4\%$ between 1930 and 1970. It rose to about $5\%$ in the 1970s and stayed there in the 1980s. It then remained around $4\%$ again between 1990 and 2020. Finally, it temporarily rose to $6\%$ during the pandemic. Overall, in the United States, the FERU is stable over time, generally around $4\%$.

\paragraph{Unemployment gap} Finally, over 1930--2023, the unemployment gap averages $+2.4$pp (figure~\ref{f:gap}). The unemployment gap reached $+20.9$pp---its highest level on record---in 1932Q3, during the Great Depression. The unemployment gap reached $-1.6$pp---its lowest level on record---in 1945Q1, at the end of World War 2. Another notable peak for the unemployment gap is $+6.4$pp in 2020Q2---its highest level in the postwar period, in the middle of the coronavirus pandemic. In the aftermath of the pandemic the unemployment gap fell to $-1.5$pp in 2022Q2. This is the lowest unemployment gap in the postwar period.

\section{Robustness}\label{s:robustness}

This section shows that the FERU is robust to alternative measures of unemployment, and to different calibrations of the cost of unemployment, cost of recruiting, and shape of the Beveridge curve.

\subsection{Alternative measures of labor force and unemployment}\label{s:u4u5}

\paragraph{Standard definition of labor force and unemployment} By definition, the labor force comprises people able, willing, and seeking to work. This definition seems natural. Anyone willing to work would also be seeking to work, otherwise they could never get a job; seeking a job reveals willingness to work. Empirically, the challenge is to determine who is seeking a job. People search with different intensity and methods. Ideally anyone searching in any way would be counted in the labor force. However, the standard US statistics only include into the labor force unemployed people who have been actively searching in the past 4 weeks. There are workers who have been searching for a job in the past year but not in the past month who are excluded from the labor force statistics, although in theory they belong there. When we apply the FERU formula to the US economy in section~\ref{s:us}, we stick to the official definition of labor force and unemployment. Here we recompute the FERU using broader definitions of labor force and unemployment---replacing the unemployment rate U3 by the broader unemployment rates U4 and U5, and adjusting the size of the labor force accordingly. To clarify that our baseline measures of $u$ and $v$ are based on the concept U3 of unemployment, we denote them by $u3$ and $v3$ here.

\begin{figure}[p]
\subcaptionbox{Alternative FERUs\label{f:ustar345}}{\includegraphics[scale=\wscale,page=23]{\pdf}}\\
\subcaptionbox{Alternative unemployment gaps\label{f:gap345}}{\includegraphics[scale=\wscale,page=24]{\pdf}}
\caption{Alternative measures of full employment in the United States, 1994--2023}
\note{The FERUs are given by $u3^* = \sqrt{u3 \times v3}$, $u4^* = \sqrt{u4 \times v4}$, and $u5^* = \sqrt{u5 \times v5}$. The unemployment rate $u3$ is measured by the \citet{UNRATE}; the unemployment rate $u4$ is measured by the \citet{U4RATE}; the unemployment rate $u5$ is measured by the \citet{U5RATE}. The vacancy level comes from figure~\ref{f:uv}. The vacancy rate $v3$ is the vacancy level divided by the number of labor-force participants, which is measured by the \citet{CLF16OV}. The vacancy rate $v4$ is the vacancy level divided by the number of labor-force participants and discouraged workers, both measured by the \citet{CLF16OV,LNU05026645}. The vacancy rate $v5$ is the vacancy level divided by the number of labor-force participants and marginally attached workers, both measured by the \citet{CLF16OV,LNU05026642}. Unemployment and vacancy rates are quarterly averages of monthly series. The gray areas are NBER-dated recessions.}
\label{f:feru345}\end{figure}

\paragraph{Broader definitions of unemployment: U4} The unemployment concept U4 includes all the workers in the standard unemployment concept U3, plus workers who want a job, are available to start a job now, have been actively searching for a job in the past 12 months, but have not been searching in the past 4 weeks because they became discouraged about their job prospects \citep{BLS23}. When asked why they did not look for work during the last 4 weeks, these workers respond for instance that ``There are no jobs available,'' or ``They have been unable to find work in the past.'' These additional workers are labelled ``discouraged workers.'' They are not classified as part of U3 because they did not actively search for work in the last 4 weeks. 

\paragraph{Broader definitions of unemployment: U5} The unemployment concept U5 includes all the workers in U4 plus workers who want a job, are available to start a job now, have been actively searching for a job in the past 12 months, but have not been searching in the past 4 weeks for other reasons than discouragement about their job prospects \citep{BLS23}. When asked why they did not look for work during the last 4 weeks, these workers respond for instance that they could not search because of family responsibilities, childcare problems, or ill health. These additional workers are not classified in U3 because they did not actively search for work in the last 4 weeks; they are not classified in U4 because they were not discouraged about their job prospects. Together with the discouraged workers, these workers compose the ``marginally attached workers.''

\paragraph{Broader definitions of the labor force} To be consistent with the definitions U4 and U5 of the unemployment level, the definition of the labor force is also adjusted approximately. These broader labor-force sizes are used to compute unemployment and vacancy rates. 

\paragraph{Broader unemployment and vacancy rates}  The unemployment rate $u4$ is the unemployment level U4 divided by an extended labor force, composed of the standard labor force plus the discouraged workers. This unemployment rate is constructed by the \citet{U4RATE}. We also compute a vacancy rate $v4$ by taking the vacancy level constructed in section~\ref{s:complete} and dividing it by the number of workers in the standard labor force plus the number of discouraged workers, both constructed by the \citet{CLF16OV,LNU05026645}. In that way, the rates $u4$ and $v4$ have the same denominator. Similarly, the unemployment rate $u5$ is the unemployment level U5 divided by an extended labor force, composed of the standard labor force plus the marginally attached workers. This unemployment rate is constructed by the \citet{U5RATE}. We compute a vacancy rate $v5$ by taking the vacancy level constructed in section~\ref{s:complete} and dividing it by the standard labor force plus the marginally attached workers, both constructed by the \citet{CLF16OV,LNU05026642}. The unemployment levels U4 and U5 were only introduced in 1994, so we can only measure $u4$, $v4$, $u5$, $v5$ for 1994--2023. Over the 1994--2023 period, the standard unemployment rate, $u3$, averages $5.6\%$. By comparison, the average value of $u4$ is $6.0\%$, and the average value of $u5$ is $6.7\%$. So the discouraged workers make up less than half a percentage point of the labor force, and the marginally attached workers make up about one percentage point of the labor force. By construction, all the vacancy rates are quite close, averaging $3.3\%$ over the period.

\paragraph{Broader definitions of the FERU} Using these broader measures of unemployment and the labor force, we construct broader measures of the FERU: $u4^* = \sqrt{u4 \times v4}$, and $u5^* = \sqrt{u5 \times v5}$ (figure~\ref{f:ustar345}). We compare these measures to the standard value of the FERU:  $u3^* = \sqrt{u3 \times v3}$. Over the 1994--2023 period, the average value of $u3^*$ is 4.2\%, the average value of $u4^*$ is 4.3\%, and the average value of $u5^*$ is 4.5\%. So the three measures of the FERU are close to each other---much closer in fact that the three measures of unemployment. The average distance between $u3$ and $u4$ is 0.4pp while the average distance between $u3^*$ and $u4^*$ is only 0.1pp; the average distance between $u3$ and $u5$ is 1.1pp while the average distance between $u3^*$ and $u5^*$ is only 0.3pp. We also see that all measures of the FERU follow the exact same patterns: the largest distance between $u3^*$ and $u4^*$ is only 0.2pp and the largest distance between $u3^*$ and $u5^*$ is only 0.6pp. The final readings of the FERU in 2023Q4 are $u3^* = 4.5\%$, $u4^* = 4.6\%$, and $u5^* = 5.0\%$. 

\paragraph{Broader definitions of the unemployment gap} Using these broader measures of unemployment and the labor force, we also construct broader measures of the unemployment gap: $u4 - u4^*$ and  $u5 - u5^*$ (figure~\ref{f:gap345}). The different measures of unemployment gap all move together. The unemployment gaps constructed with the broader measures of unemployment are larger than the baseline unemployment gap, because the unemployment rates $u4$ and $u5$ are larger than $u3$, but the differences are not as large as the differences in unemployment rates because part of it is absorbed by the differences in FERUs. Over the 1994--2023 period, the gap $u3 - u3^*$ averages $1.5$pp, the gap $u4 - u4^*$ averages $1.7$pp, and the gap $u5 - u5^*$ averages $2.2$pp. Since the unemployment gap is larger with the broader measures of unemployment, the economy appears at full employment---and not inefficiently tight---in 2019 under U5. The final readings of the unemployment gap in 2023Q4 are $u3 - u3^* = -0.7$pp, $u4 - u4^* -0.6$pp, and $u5 - u5^* = -0.4$pp. So all unemployment gaps are negative, which indicates that the labor market is inefficiently tight by all measures of unemployment at the end of 2023.

\subsection{Alternative calibrations of the cost of unemployment, cost of recruiting, and elasticity of the Beveridge curve}\label{s:ms21}

\paragraph{Generalized FERU formula} We derive the formula $u^* = \sqrt{uv}$ by assuming that jobseekers do not contribute to social welfare, that a 1\% increase in unemployment affords a 1\% decrease in vacancies (Beveridge elasticity of 1), and that one vacancy requires one recruiter. These assumptions are based on evidence for the United States. It is possible to generalize the formula for the FERU under a more general calibration. If the Beveridge elasticity is $\e \neq 1$, the social value of nonwork is $\z>0$, and the recruiting cost is $\k\neq 1$, then the FERU becomes
\begin{equation}
u^* = \bp{\frac{\k \e}{1-\z}\cdot\frac{v}{u^{-\e}}}^{1/(1+\e)},
\label{e:MS21}\end{equation}
as showed by \citet[proposition 3]{MS21}. Note that by setting $\e=1$, $\k=1$, and $\z=0$ in formula \eqref{e:MS21}, we recover the simpler formula $u^* = \sqrt{uv}$.

\begin{figure}[t]
\includegraphics[scale=\wscale,page=7]{\pdf}
\caption{FERU under alternative calibration of the social costs and benefits of unemployment}
\note{The solid, pink line reproduces the FERU from figure~\ref{f:ustar1951}.  The dotted, purple line is the FERU under the Beveridge elasticity $\e$ estimated by \citet[figure 6]{MS21}, the social value of nonwork $\z=0.26$, and the recruiting cost $\k=0.92$, as constructed by \citet[figure 7B]{MS21}. The gray areas are NBER-dated recessions.}
\label{f:MS21}\end{figure}

\paragraph{Distance between basic and generalized FERU} The generalized formula requires to keep track of three statistics in addition to the unemployment and vacancy rates ($\e$, $\z$, $\k$), so it is harder to compute than $u^* = \sqrt{uv}$. The generalized formula is especially difficult to use in real time because it requires to keep track of the slope of the Beveridge curve, which is hard to do when the curve shifts. By setting the statistics to reasonable but fixed values, we obtain a formula that is simpler and more user-friendly, and therefore appropriate to measure full employment in real time. In the United States, however, the two formulas yield almost identical unemployment rates (figure~\ref{f:MS21}). The generalized FERU formula is applied using under the Beveridge elasticity $\e$ estimated by \citet[figure 6]{MS21}, the social value of nonwork $\z=0.26$, and the recruiting cost $\k=0.92$, as constructed by \citet[figure 7B]{MS21}. Between 1951 and 2019 the two FERUs only depart by $0.2$pp on average; they never depart by more than 0.5pp.

\section{Explaining deviations from full employment}\label{s:explaining}

Despite the US government's full-employment mandate, the US labor market generally deviates from full employment (figure~\ref{f:state}). Here we discuss why the US labor market has consistently fallen short of full employment in the past century.

\subsection{Great Depression and its aftermath} 

During the Great Depression and its aftermath, the US economy was exceedingly slack. From the beginning of 1930, when our data begin, to the end of 1941, when the United States enter World War 2, the unemployment gap averages $+9.6$pp (figure~\ref{f:feru1930}). So the US economy was extremely far from full employment. One factor that might have played a role is that the US government and the Federal Reserve did not have a full-employment mandate at the time. The mandate was introduced with the Employment Act of 1946, as a result of the Great Depression. A second factor is that the Federal Reserve was committed to the gold standard, which created a deep deflation in the early 1930s, with dramatic consequences \citep{ET00}. A third factor is that the Fed failed to curb recurrent banking panics in the 1930s \citep[chapter 7]{FS63}. Overall, as former Fed chairman \citet[p. xvii]{B22} writes, ``Blaming the Depression entirely on the Fed is an exaggeration, but the relatively new and unseasoned central bank did perform poorly.''

\subsection{World War 2, Korean War, Vietnam War} 

\paragraph{Inefficient tightness in wartime} The US labor market was pulled out of its Great Depression slackness by World War 2. In fact, the labor market became inefficiently tight during the war, with tightness averaging $3.2>1$ over the 1942--1945 period (figure~\ref{f:tightness1930}). The labor market was once again inefficiently tight during the Korean War, with tightness averaging $1.2>1$ over the 1951--1953 period, and during the Vietnam War, with tightness averaging $1.2>1$ over the 1966--1969 period (figure~\ref{f:tightness1951}). 

\paragraph{Source of tightness} Why was tightness so high during the wars? Part of the reason is that the government spends a lot during wars, and the amount of spending during these three major wars was extremely large \citep{RS98}. Such expenditure boosts aggregate demand and then increase tightness \citep[figure 2]{MS19}. Another part of the reason, is that millions of potential labor-force participants were sent abroad on military duty \citep{VA23}. Such drastic reduction in labor force reduces labor supply, which raises labor-market tightness and reduces the unemployment rate among the workers who stayed in the United States \citep[figure 4]{MS22}.

\paragraph{Absence of tightening by the Fed: World War 2} So why didn't the Fed tighten monetary policy to reduce tightness in wartime? Indeed, a high real interest rate curbs aggregate demand, which reduces tightness and raises unemployment \citep[figure 5]{MS22}. An appropriate increase in interest rates could have brought tightness back to its full-employment level of 1 \citep[figure 7]{MS22}. In the case of World War 2, there is a simple answer. As \citet[p. xviii]{B22} explains, during and shortly after World War 2, ``at the Treasury's request, the Fed held interest rates at low levels to reduce the government's cost of financing the war.'' 

\paragraph{Absence of tightening by the Fed: Korean War} The same happened at the beginning of the Korean war, when ``facing new hostilities in Korea, President Truman pressed the Fed to keep rates low'' \citep[p. xviii]{B22}. The Fed did rebel and was allowed to phase out the low interest-rate peg that had been in place. But the phasing out came too late to cool down the Korean-War labor market. 

\paragraph{Absence of tightening by the Fed: Vietnam War} The situation during the Vietnam War was different \citep[pp. 20--22]{B22}. The Fed raised interest rates by half a percentage point at the end of 1965, at the exact time when the economy had reached full employment (figure~\ref{f:gap1951}). However, President Johnson was furious that the Fed tightened monetary policy. He needed a low rates to help finance the Vietnam War. Despite the pressure exerted by Johnson, the Fed continued increasing rates in 1966, which rapidly cooled the labor market (figure~\ref{f:gap1951}). Worried about a possible recession, the Fed reversed its previous tightening. Under pressure from the White House, and facing a chaotic political situation, the Fed continued swinging between tightening and loosening until 1970. The absence of decisive tightening explains why the labor market was so hot in 1966--1969.

\subsection{Postwar period} 

\paragraph{Inefficient slackness in the postwar period}  In the postwar period, the US labor market was generally inefficiently slack (figure~\ref{f:gap}). A manifestation of such pervasive slack is that the unemployment gap averaged 1.6pp between 1946 and 2019. Another manifestation is that the labor market was not at full employment once from 1970Q1 to 2018Q1: it was inefficiently slack for almost half a century.

\paragraph{Elevated target} A first possible reason to explain this slackness is that the Fed and other policymakers often use the NRU computed by the CBO to measure full employment. Over 1949--2019, the NRU averages 5.5\% \citep{NROU}. This is 1.4pp above the average of the FERU between 1949 and 2019, which is 4.1\%. So policymakers might have targeted an unemployment rate that was just too high (figure~\ref{f:nru}). The average distance between the FERU and the NRU by itself explains almost the entire average postwar unemployment gap. 

\paragraph{Another elevated target} Another measure of full employment that policymakers sometimes use is the NAIRU---although there is no standardized time series for it. Just like the NRU, the NAIRU appears to be significantly higher than the FERU. For instance, the NAIRU computed by \citet[figure~8B]{CEG19} using state-of-the-art techniques averages 5.7\% between 1960 and 2018. This is 1.4pp more than the average FERU between 1960 and 2018, which is 4.3\%. Once again, by using the NAIRU, policymakers would have targeted an unemployment rate that was just too high (figure~\ref{f:nru}). 

\begin{figure}[t]
\includegraphics[scale=\wscale,page=25]{\pdf}
\caption{FERU, NRU, and NAIRU in the United States, 1930--2023}
\note{The FERU comes from figure~\ref{f:ustar}. The NRU is constructed by the \citet{NROU}. The short-term NRU is constructed by the \citet{NROUST}. The NAIRU is constructed by \citet[figure~8B]{CEG19}. The gray areas are NBER-dated recessions.}
\label{f:nru}\end{figure}

\paragraph{Priority to inflation} A second reason that might explain the slackness of the US labor market in the postwar period, especially after 1970, is that the Fed prioritized inflation at the expense of unemployment.  \citet{T11a} reviews policy directives by the Federal Open Market Committee (FOMC) and finds that it made no reference to unemployment or full employment between 1979 and 2008---despite the dual mandate introduced in 1977. Instead, \citeauthor{T11a} finds that the FOMC preferred ``to state its objectives in terms of price stability and economic growth.'' This changed at the end of 2008, when the FOMC started mentioning its dual objective of ``maximum employment and price stability'' in policy directives and statements. \citet{KGK19} also detect this focus on inflation in FOMC transcripts. They find that from 1960 to 2010 FOMC discussions increasingly emphasized inflation relative to unemployment, and that this shift occurred during the Volcker era and continued even as inflation declined. They conclude that ``the emphasis on inflation has become entrenched and disconnected from actual inflation.''

\paragraph{Possible reasons behind the prioritization of inflation} This focus on inflation might be due to a change in the Fed's preferences or in macroeconomic theory. But partly it also appears to come from Congress. \citet{HS16} examine legislative activity to determine when Congress pressures the Fed, and whether this pressure affects monetary policy. They find that by the late 1980s Congress shifted from threatening the Fed when unemployment was high to threatening when inflation was high. This finding is consistent with \citet[p. 377]{W87a}'s view that ``By the mid-1980s full employment had been all but erased as a major political issue in the United States.'' In fact, \citet[p. 395]{W87a} argues that although the Kennedy CEA identified an unemployment rate of 4\% as full employment, in the following decades ``more conservative economists [offered] ever-increasing rates of unemployment as the `true' definition of full employment.'' 

\subsection{Great Recession} 

\paragraph{Zero lower bound} The Great Recession saw the highest unemployment gap of the 1946--2019 period, and it presented new challenges to the Fed (figure~\ref{f:gap}). Although the unemployment gap skyrocketed in 2008--2009, the Fed was unable to respond because it ran against the zero lower bound on nominal interest rates from the end of 2008 until 2015 \citep{FEDFUNDS}. The Fed could not stimulate aggregate demand through lower interest rates because it was constrained by the zero lower bound, so it could not boost tightness and lower unemployment \citep[figure 8]{MS22}. Hence, unemployment remained inefficiently high until the end of 2018.

\paragraph{Elevated target} The Fed did resort to unconventional monetary policy, such as forward guidance and quantitative easing, to reduce long-term interest rates \citep{K18a}. But the effectiveness of such policies, both empirically and theoretically, is debatable \citep{GHH18,MS21a}. Moreover, the Fed may not have used these policies aggressively enough because once again the unemployment rate that they targeted was too high. The Fed commonly uses the CBO's NRU to indicate full employment. During the Great Recession the CBO adjusted the NRU upward by 0.9pp because they believed that structural factors temporarily kept the unemployment rate high. As a result, in 2011Q4, the short-term NRU reached 5.8\% (figure~\ref{f:nru}). We do find that the outward shift of the Beveridge curve after the Great Recession lead to an increase in the FERU by 0.5pp, but the FERU only stood at 4.5\% in 2011Q4---1.3pp below the short-term NRU. 

\subsection{Coronavirus pandemic} 

\paragraph{Zero lower bound} The coronavirus pandemic lead to a sharp slowdown in economic activity. In 2020, the US economy reached the largest unemployment gap since the Great Depression, at $+6.4$pp (figure~\ref{f:gap2020}). As during the Great Recession, the Fed was unable to respond more aggressively to the slackness of the economy because it rapidly hit the zero lower bound \citep{FEDFUNDS}.

\paragraph{Recovery} The US economy recovered fairly rapidly from the pandemic, however, thanks to aggressive expansionary fiscal policy \citep{R21}. The US economy reached full employment in 2021Q2, and continued tightening after that. In 2022Q2, labor-market tightness reached 1.97, a level it had not seen since the end of World War 2 (figure~\ref{f:tightness}). It is only then, in spring 2022, that the Fed started tightening monetary policy \citep{FEDFUNDS}. It is unclear why the Fed start tightening monetary policy earlier. One entire year passed between the labor market becoming too tight (spring 2021) and the Fed increasing rates (spring 2022). This delay is especially surprising since inflation was also above its target of 2\% at the time: core inflation was 3.7\% in 2021Q2 and rose to 6.3\% in 2022Q1 \citep{CPILFESL}. This delay, combined with the two years monetary policy takes to be fully effective \citep{C12}, explains why the labor market was inefficiently tight until the end of 2023.

\section{Conclusion}

To conclude, we discuss possible uses of the FERU formula for monetary policy.

\paragraph{Full-employment target} The Federal Reserve is mandated to stabilize the economy at full employment. However, there is no agreed-upon standard of what is the unemployment rate at full employment, which makes it difficult for the Fed to design policy to achieve full employment, and for outside observers to assess the performance of the Fed \citep[p. 3]{D77}. We have argued that the legal notion of full employment should be translated into the economic concept of social efficiency, and accordingly, that the unemployment rate $u^* = \sqrt{uv}$ is the appropriate measure of full employment. The Fed could use $u^*$ as their full-employment target---especially since $\sqrt{uv}$ can be measured in real time. 

\paragraph{Optimality of the full-employment mandate} We have argued that full employment and social efficiency are one and the same. This is a convenient property: since the Fed and social planner have the same objective, insights from optimal policy analysis have direct practical implications. In particular, maintaining the economy at full employment is the optimal policy in a range of monetary models with a Beveridge curve in which the Fed, by setting interest rates, can control aggregate demand and thus the unemployment rate. 

\paragraph{Optimal monetary policy with fixed inflation} For instance, the model by \citet{MS22} has a horizontal Phillips curve, so inflation is fixed and the Fed does not need to worry about its price-stability mandate. Then the optimal monetary policy is to keep the economy at full employment: the unemployment rate should always be at $u^* = \sqrt{uv}$.

\paragraph{Optimal monetary policy with divine coincidence} The model by \citet{MS24} has a downward-sloping Phillips curve: higher unemployment leads to lower inflation. Furthermore, the Phillips curve goes through the point of divine coincidence: when the unemployment rate is efficient, inflation is on target. Therefore there is no trade-off between inflation and unemployment. Maintaining the unemployment rate at its efficient level also maintains inflation on target. In that model the optimal monetary policy is once again to maintain the unemployment rate at $u^* = \sqrt{uv}$.

\paragraph{Optimal monetary policy without divine coincidence}  In models in which the divine coincidence fails, monetary policy faces a tradeoff between closing the unemployment gap and bringing inflation to its target. It is no longer optimal to eliminate the unemployment gap by targeting the FERU. The unemployment gap remains useful to design policy, but it has to be weighted against the inflation gap to determine optimal monetary policy.\footnote{The same logic applies to optimal fiscal policy. It is not optimal to close the unemployment gap with government spending because government spending is not a perfect substitute for private spending. Nevertheless, the unemployment gap is a key statistic to determine the optimal level of government spending---together with the elasticity of substitution between public and private consumption \citep{MS19}.}

\bibliography{\bib}
\newpage
\appendix

\section{Theoretical underpinning of existing full-employment criteria}\label{s:underpinning}

Since World War 2, various criteria for full employment have been proposed. Our analysis provides a theoretical underpinning for several of these ideas.

\subsection{Beveridge's criterion} 

\citet[p. 18]{B44} states that ``Full employment\dots means having always more vacant jobs than unemployed men, not slightly fewer jobs.'' \citet[p. 39, chart 5]{R57} applies Beveridge's verbal criterion to the United States. He computes the ratio between number of vacancies and number of unemployed workers and examines when the ratio crosses 1. Our theory provide a foundation for Beveridge's idea that full employment can be assessed by comparing the numbers of vacant jobs and unemployed workers. We find that in the United States, the labor market is at full employment when jobseekers and vacancies are in equal number. Unlike Beveridge, we find that having more vacancies than jobseekers is a sign of inefficient tightness. Beveridge thought that the labor market could be either too slack, when $v/u<1$ or at full employment, when $v/u\geq 1$. In contrast, we argue that full employment is achieved exactly when $v/u=1$, and that on either side of full employment, the labor market can be too slack, when $v/u<1$, or too tight, when $v/u>1$.

\subsection{Council of Economic Advisers' criterion} 

In the early postwar period, from 1946 to 1956, the Council of Economic Advisers (CEA) used an unemployment rate of 3\% as a marker of full employment \citep[p. 8]{D77}. The CEA's criterion seems appropriate, since the FERU was as low as 3.0\% in 1953 and averaged 3.5\% over 1946--1956 (figure~\ref{f:ustar}). Then the CEA started raising their unemployment rate at full employment---just as the FERU was rising. In 1962, the CEA announced that an unemployment rate of 4\% was ``a reasonable and prudent full employment target for stabilization policy'' \citep[p. 10]{D77}. Coincidentally, our FERU is 4.0\% at the end of 1961. Then, in 1969, \citet[p. 280]{B69} reported that ``Since the [CEA] identified an unemployment rate of 4\% with a condition of practically full employment, this figure served as a constant in the equation for computing the potential output.'' Again, this is almost the same as the average value of our FERU over 1962--1969, which is 3.9\%. Then, facing rising unemployment in the 1970s, the CEA moved away from a numerical target for full employment. When testifying in front of Congress in 1975, Greenspan, who was then chairing the CEA, was asked what the target for full employment was. He responded: ``I do not think we should set a target'' \citep[p. 13]{D77}.

\subsection{Burns's criterion}

Just before he became chairman of the Fed, and after his stint as chairman of the CEA, \citet[p. 284]{B69} argued that the number of 4\% used by the CEA at the time was not compelling because it does not incorporate information on vacancies. He wrote that ``we need to develop comprehensive data on job vacancies, so that it will no longer be necessary to guess whether or when a deficiency in aggregate demand exists.'' Thirty years after Burns's article, the government started collecting data on vacancies as part of the JOLTS, which started in 2001. In this paper, we combine statistics on vacancies and unemployment to compute the unemployment rate  $u^*$ consistent with full employment. By computing the unemployment gap $u-u^*$, we can then assess in real time whether ``a deficiency in aggregate demand exists,'' or whether an excess of aggregate demand exists. Indeed, a positive unemployment gap indicates that aggregate demand is insufficient to sustain full employment; by contrast, a negative unemployment gap indicates that aggregate demand is excessive and should be restricted to attain full employment \citep[figures 5--7]{MS22}.

\subsection{Full Employment and Balanced Growth Act's criterion} 

Although it does not define what full employment is, the Full Employment and Balanced Growth Act does give a numerical target for full employment: an unemployment rate of not more than 4\% by 1983 \citep[p. 1893]{FEBGA}. This target was quite ambitious. It is lower than the FERU that we compute in 1983, $u^*=5.0\%$, lower also than the FERU in 1978, $u^* = 5.3\%$, and lower than the average value of the FERU throughout the 1970s, which we estimate at 4.7\%. But the number of 4\% was in line with the unemployment rate of 4\% that was used by the CEA as full-employment target throughout the 1960s, and in fact with the average value of the FERU throughout the 1960s, which we measure at $3.9\%$. So the numerical target adopted by the Full Employment and Balanced Growth Act was in line with average value of the FERU in the decade before the law was enacted.

\subsection{Powell's criterion} 

During the press conference following the May 2022 FOMC meeting, journalist Howard Schneider asked Fed Chair Jerome Powell which tightness the Fed might target. \citet[pp. 12--13]{P22} responded: ``So in terms of the vacancy-to-unemployment ratio, we don't have a goal in mind.\dots I think when we got to one-to-one in the, you know, in the late teens, we thought that was a pretty good number. But, again, we're not shooting for any particular number.'' Our analysis confirms that a vacancy-to-unemployment ratio of 1 is what the Fed should target to satisfy its full-employment mandate. The Fed should be shooting for that particular number.

\end{document}